\documentclass{article}

\usepackage{arxiv}

\usepackage[utf8]{inputenc} 
\usepackage[T1]{fontenc}    
\usepackage{hyperref}       
\usepackage{url}            
\usepackage{booktabs}       
\usepackage{amsfonts}       
\usepackage{nicefrac}       
\usepackage{microtype}      
\usepackage{algorithm}
\usepackage{algorithmic}
\usepackage{lipsum}
\usepackage{graphicx}
\graphicspath{ {./images/} }

\title{A Novel Privacy-Preserved Recommender System Framework based on Federated Learning}

\author{
 Jiangcheng Qin* \\
  College of Information Science and Engineering\\
  Ningbo University\\
  Zhejiang, China \\
  \texttt{qjc@nbu.edu} \\
   \And
 Baisong Liu \\
  College of Information Science and Engineering\\
  Ningbo University\\
  Zhejiang, China \\
  \texttt{lbs@nbu.edu} \\
  \And
}

\begin{document}
\maketitle
\begin{abstract}
Recommender System (RS) is currently an effective way to solve information overload. To meet users' next click behavior, RS needs to collect users' personal information and behavior to achieve a comprehensive and profound user preference perception. However, these centrally collected data are privacy-sensitive, and any leakage may cause severe problems to both users and service providers. This paper proposed a novel privacy-preserved recommender system framework (PPRSF), through the application of federated learning paradigm, to enable the recommendation algorithm to be trained and carry out inference without centrally collecting users' private data. The PPRSF not only able to reduces the privacy leakage risk, satisfies legal and regulatory requirements but also allows various recommendation algorithms to be applied.
\end{abstract}


\section{Introduction}
With the explosive development of online services, products, and information resources, the phenomenon of "information overload" is becoming more serious, which brings a significant information burden to people. Recommender Systems (RS) is currently an effective way to solve information overload and realize personalized information services. Currently, RS has played a core role in many domains, such as e-commerce (Alibaba, Amazon), video service (Netflix, YouTube, and TikTok), online-music services (Pandora, Yahoo), Advertising (Google, Facebook), both the user and service providers benefit from the personalized RS. However, the current RS exists many potential risks, in which privacy leakage is one of the top concerns \cite{bobadilla2013recommender}.
\begin{figure}[b]
    \centering
    \includegraphics[width=0.65\textwidth]{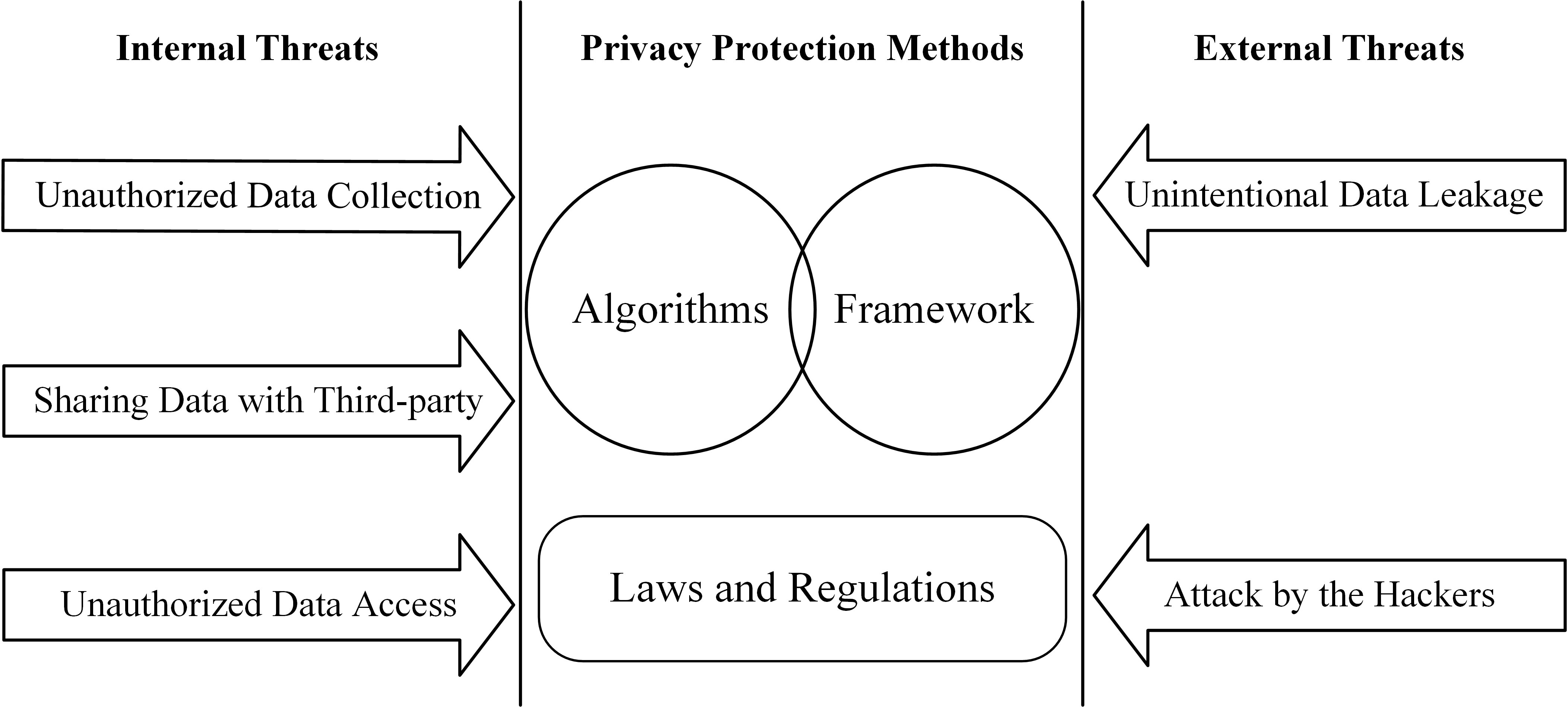}
    \caption{Internal and External Privacy Threats to the RS and Available Countermeasures}
    \label{fig:1}
\end{figure}

The ultimate goal of RS is to understand user preference better than users themselves \cite{kelly2003implicit}\cite{zhao2016user}. RS needs to collect users' personal information and behavior to achieve a comprehensive and profound user preference perception to meet users' next click behavior. The prediction quality varies depending on the scale, diversity, and timeliness of the data collected. However, these data quality factors are also positively related to privacy risks and the degree of harm caused by leaking these data. This is an unavoidable "privacy-personalization trade-off problem \cite{ramakrishnan2001being}\cite{zhang2019proactive}"to the current RS. The RS's privacy risk is mainly due to internal and external improper operations and attacks on the centrally collected and stored user information. As illustrated in Figure 1, internal privacy threats come from a curious or malicious service provider. Driven by commercial interests, service providers may violate privacy contracts and collect user data without authorization or share data with third parties. Internal staff may also use access permissions to spy on users' privacy. External privacy threats mainly come from hacker attacks. Due to the lack of adequate data security protection, data is maliciously stolen. Sometimes there will be unconscious data leakages, such as data surveillance by law enforcement agencies or publicly anonymous data being de-anonymized (de-anonymization attack \cite{narayanan2008robust}) by a third party. 

In May 2015, eBay was attacked by hackers, resulting in the disclosure of 145 million user accounts, including user names, addresses, data-of-birth, and account password. In March 2018, Facebook leaked tens of millions of users' personal information. Many Internet companies have similar incidents, leading to large-scale user privacy data leakage, triggering user trust crisis. Consequently, people raised their attention to protect their data privacy and security and avoid their private information being collected by Internet applications. Relevant data security and privacy protection laws and regulations have been promulgated as well. The US proposed the "California Consumer Privacy Act " (CCPA)\footnote{https://oag.ca.gov/privacy/ccpa}, the EU published "General Data Protection Regulation" (GDPR)\footnote{https://gdpr-info.eu/}, Singapore published "Personal Data Protection Act" (PDPA)\footnote{https://www.pwc.com/sg/en/personal-data-protection.html}, and "Data Security Management Measures (Draft for Comment)" published by China. These regulations rely on a fundamental principle: \textit{It is up to individuals, groups, or organizations to decide when, how, and to what extent to communicate information about them to others. By Alan F. Westin} \cite{westin2003social}. These regulations have made user data collection face many obstacles without users' full authorization. In this context, the user's local device firmly locks the critical data, forming the isolated data islands that are unable to aggregate, cause the RS to no longer personalized, and user experience also declines. 

The emergence of Federated Learning (FL) connects isolated data under the premise of ensuring data security and user privacy, thereby providing the possibility to train accurate and robust machine learning models. McMahan et al.\cite{mcmahan2017communication} first introduced the term FL in 2016 and applied it to update smart-phones' voice recognition and text entry models. The Google team utilize the user's mobile phone to train the local user interaction data, which can continuously optimize the user's local language prediction model, and through the "Federated Averaging" algorithm (FedAvg), by averaging all stochastic gradient descent (SGD) result on user's terminal with a central server, a globally optimized model is obtained. Currently, the study of FL has been carried out in many application fields \cite{tran2019federated}\cite{li2020federated}\cite{brisimi2018federated} related to privacy protection and data security. 

In this paper, we introduced the FL paradigm into the RS and proposed a novel privacy-preserved RS framework based on FL (PPRSF) that enables the training and online inference under the premise of preserving user privacy, satisfying legal and regulatory requirements. 

\section{Related Work}
In this section, we review relevant studies of privacy-preserved RS and federated learning to position our research in relation to existing studies.
\subsection{Privacy-preserved RS}
In response to the privacy threats, privacy-preserved RS use specially designed architectures or algorithms as countermeasures.

\paragraph{Architecture-based Solutions}
Architecture-based solutions are designed to minimize the threat of data leakage. For example, using distributed storage of user data can effectively reduce the damage caused by a single data source exposure, or distributed the recommendation process to increase the difficulty of unauthorized access to data. Based on distributed data storage, Heitmann et al.\cite{heitmann2010architecture} proposed a candidate architecture, in which users hold local configuration data, and the users decide which data can be disclosed to which service provider, only holding specific certificates the application can access this part of configuration data through API and use it for recommendation calculation. This method allows users to make data disclosure decisions, but it cannot avoid collecting open data by curious service providers, while users often choose to open data to obtain better services. Based on the distributed recommendation process, some studies have proposed to design and implement the recommendation process in a pure P2P system \cite{mordacchini2010p2p}\cite{hecht2012radiommender}\cite{han2004scalable}, compare local user data with another user for similarity to obtain possible recommendation results, remove the role of the central server to avoid individuals Centralized storage of information. However, such methods may still leak user data to other users and impose requirements on local devices' computing power.

\paragraph{Algorithms-based Solutions}
The algorithm-based solution makes certain modifications to the original data so that even if the data or model output is obtained by a third party, sensitive information will not be exposed. The main methods including differential privacy (DP) and encryption algorithms.

\begin{itemize}
    \item DP is a privacy model based on data fuzzification, which can alleviate the risk of inferring users' private data through the output of the recommendation system. The core idea of DP is that when the enemy tries to query individual information from the database, the data is confused by noise, so that the enemy cannot distinguish the sensitivity of the individual from the result. McSherry and Mironov \cite{mcsherry2007mechanism} first applied DP to collaborative filtering, using the Laplace mechanism to generate corresponding noise on the input score, which can shield the impact of the change of a single record on the final result. Follow-up research includes\cite{friedman2016differential}\cite{berlioz2015applying} However, due to noise interference, DP cannot guarantee privacy and the accuracy of recommendations at the same time. Moreover, the continuous increase of the data received by the recommendation system is a dynamic process, and the application of DP in a continuous environment is still a problem.
    
    \item The encryption-based solution can reduce the privacy risk caused by the leakage of user data. The current main method is homomorphic encryption (HE) \cite{HE-erkin2012generating}\cite{HE-gentry2009fully}. The use of a secure multi-party calculation protocol can ensure that the user input data is in an encrypted state, but still can perform accurate calculations, and ensure that the recommended accuracy is the same as in the non-encrypted state. Its disadvantage is that it needs to consume more computing time, storage space and communication cost, and it is only suitable for offline computing.
\end{itemize}

\subsection{Federated Learning}
FL following the \emph{"minimizing the amount of collection data"} principle \cite{mini-collection-granello2004online}. Compared with traditional machine learning methods that require training data to be concentrated in one machine or a data center, federated learning uses mobile devices (clients) dispersed in the hands of thousands of users to train machine learning models collaboratively, and all training data remains in their respective devices, thus protecting the privacy of users. In this way, avoid privacy risks and data security issues caused by centralized data collection and storage, but can still utilize all users data. The relevant background knowledge can be obtained from literature reviews \cite{survey-aledhari2020federated}\cite{survey-mothukuri2020survey}\cite{survey-yang2019federated}\cite{li2020federated}.

We here present the definition of federated learning as:

\textbf{Federated Learning} \emph{ is a learning framework in which multiple participants (mobile devices, organizations) holding different data collaborate to complete a specific machine learning task.} A Host (a central server or an individual member) initiates a learning task, and under its coordination, each member trains the local data. Then, the initiator aggregates all encrypted training results safely as a global model, sends it back to the participants, and repeats the above process until the global model achieve the training objective. Finally, share the global optimized machine learning model with all participated members. A participant's raw data will not be exchanged or transferred throughout the process and ensure no participant can predict other members' original data.

More generally, assume $N$ different participants $F_i$, where $i\in[1,N]$. Each member $F_i$ holds different local raw data $D_i$. Given a machine learning task $T$:
\begin{itemize}
    \item Denote $M_i$ as the model trained by each member $F_i$ based on its local data $D_i$. 
    \item In the traditional setting, collect every member's local data together as $D_{Sum}$ and store it centrally. Denote $M_{Sum}$ as the model trained on data $D_{Sum}$.
    \item In the FL setting, aggregated every member's local model $M_i$ and form the global model $M_{FL}$.
\end{itemize}

Set $P$ as a performance measure (Accuracy, MSE, Precision, Recall, etc.). Denote $P_i$ as the performance of model $M_i$, $P_{Sum}$ as the performance of model $M_{Sum}$, and $P_{FL}$ as the performance of model $M_{FL}$. A valid FL system will have $P_{FL} >P_i$ , where $i\in [1,N]$, and its global objective is to minimize the $|P_{Sum}-P_{FL}|$, that is able to fully approximate the ideal model.

If there exist a non-negative real number $\delta$ makes the establishment of an inequality as following:
$$
|P_{Sum} - P_{FL}| < \delta
$$
It is said that the federated learning algorithm reaches $\delta$- precision loss\cite{survey-yang2019federated}. FL allows a certain degree of deviation in the performance of the training model, but provides data security and privacy protection for all participants.

Many efforts have been devoted to implementing FL based machine learning models, including Deep Neural Networks (DNN) \cite{DNN-ma2020distributed}\cite{DNN-ye2020federated}\cite{DNN-zhang2020fedmec}, Linear Regression (LR) \cite{LR-chen2018privacy}\cite{LR-nikolaenko2013privacy}\cite{LR-sanil2004privacy}, Gradient Boosted Decision Trees (GBDT) \cite{GBDT-li2020privacy}\cite{GBDT-lu2019differentially}\cite{GBDT-zhao2018inprivate}. In addition, ensemble with different privacy-preserving methods, such as Differential Privacy (DP) \cite{DP-triastcyn2019federated}\cite{DP-wei2020federated} and Homomorphic Encryption (HE) \cite{HE-hao2019efficient}\cite{HE-zhang2020batchcrypt} to prevent internal and external privacy threats \cite{threats-bhagoji2019analyzing} and attacks \cite{attack-bagdasaryan2020backdoor}. These studies provide valid theoretical basis for introducing FL into the RS. 

Ammad-ud-din et al.\cite{FCF-ammad2019federated} introduced the first FL-based Recommender System to adopt the CF method (FCF) that utilized a stochastic gradient approach. Following FCF's work, Lin et al.\cite{fedrec-lin2020fedrec} proposed a novel and generic federated recommendation model (FedRec) for rating prediction with explicit feedback. Muhammad et al. \cite{fedfast-muhammad2020fedfast} proposed FedFast for making accurate distributed recommendations using DNN and FL. They adjust the Embedding's size to ensure its robustness on devices with insufficient computing power (mobile phone). Jalalirad et al. \cite{SE-FRS-jalalirad2019simple} introduced a simple and efficient extension of FL (SE-FRS) for recommender systems inspired by REPTILE meta-learning techniques \cite{REPTILE-nichol2018first}. Moreover, they confirmed the importance of utilizing embedding technology with an empirical study. 

The FL show promising results in RS but still on the initial stage, existing studies mainly utilized feasible FL machine learning algorithms such as Stochastic Gradient Decent (SGD) \cite{FCF-ammad2019federated} or DNN \cite{DNN-qi2020privacy} to implement Collaborative Filtering based RS, and use HE and DP to optimized the data privacy protection in FL based RS \cite{privacy-dou2019privacy}. These studies lack the overall consideration of the RS architecture. Treating all user data as private data is not practical and will encounter problems (unable to initialize model, data transmission overload, the insufficient computing power of user device) in a real application scenario. In response to these problems, we proposed the a novel privacy-preserved recommender system framework (PPRSF).

\section{PPRSF}
\label{sec:headings}
This section proposed our PPRSF (illustrated as Figure 2). We first present the preliminary work that initialize the member setting and data setting. Then, introduce the framework structure, the function, and the significance of each layer. Lastly, illustrate the FL based training method of essential models, the Global Recall Model, and the Local Ranking Model.

\begin{figure}[ht]
    \centering
    \includegraphics[width=0.65\textwidth]{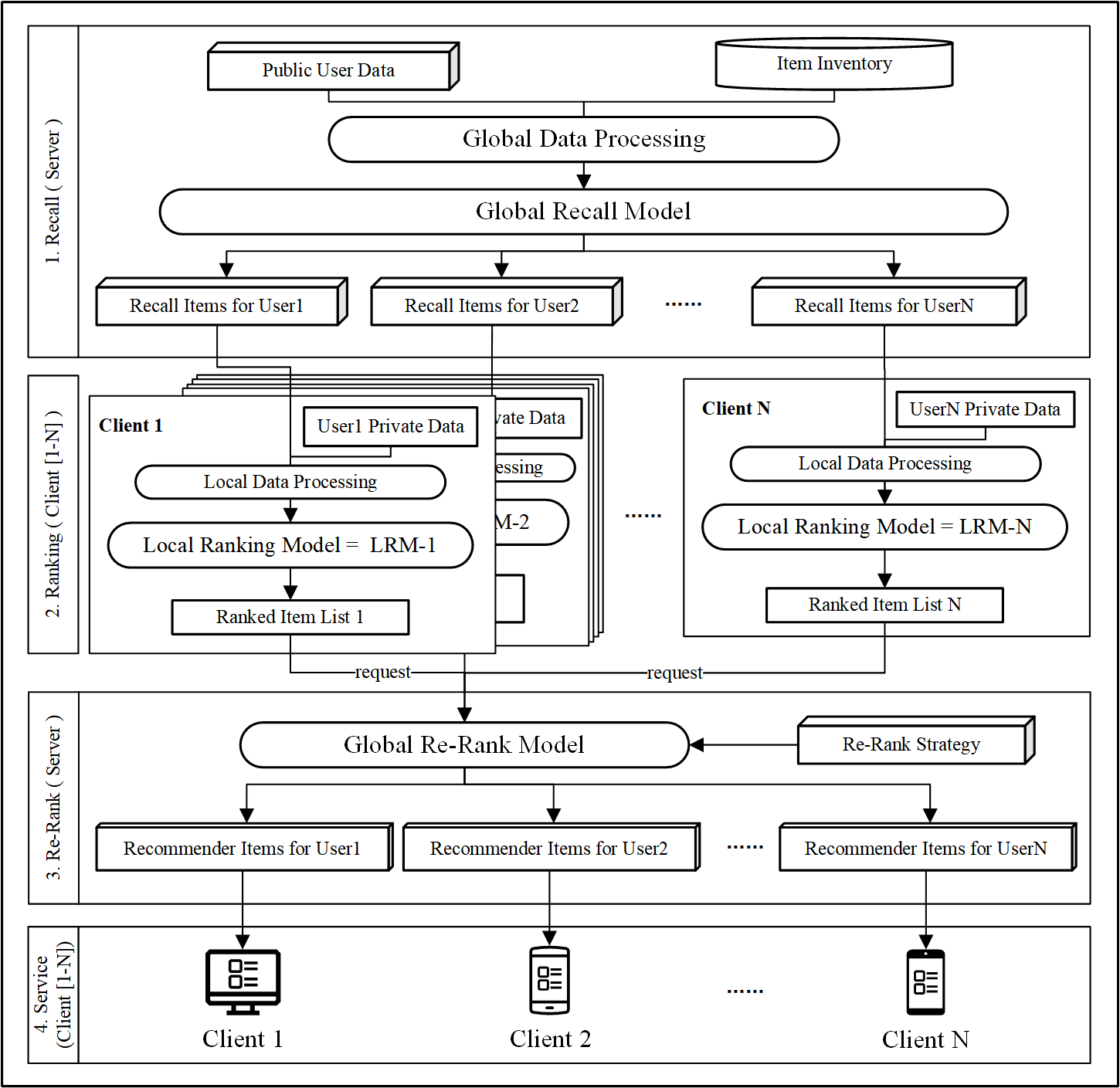}
    \caption{Privacy-preserved Recommender System based on Federated Learning Framework (PPRSF) with 4-layers Hierarchical Structure}
    \label{fig:2}
\end{figure}

\subsection{Preliminary}
The federated learning system (FLS) developed by the Google team uses numerous users' smartphones as computing entities for algorithm training. This type of FLS constructed by a central server and massive terminal devices is called Centralized Cross-device FL Architecture \cite{survey-yang2019federated} ,illustrated in Figure 3. The Cross-device FL system has a relatively large number of members (usually smartphone), and each member has small local raw data and relatively low computation power \cite{mcmahan2017communication}. Therefore, this type of FL system relies on a mighty host to manage a stable communication network and cannot perform a complex machine learning task due to the limitation of member's computation power. Generally, in the form of a centralized network. Centralized Cross-device FL Architecture allows each member to train the model locally according to their conditions, aggregate the desensitization local model parameters on the central server, form a global model, and then send it back to each member. The member updates its local model until the global model has achieved the training object. In RS, the user's computer and smartphone as the client devices, and a central server (RS server) conducts data collection, model training, and online inference. The Centralized Cross-device FL architecture applies to the RS scenario.

\begin{figure}[ht]
    \centering
    \includegraphics[width=0.55\textwidth]{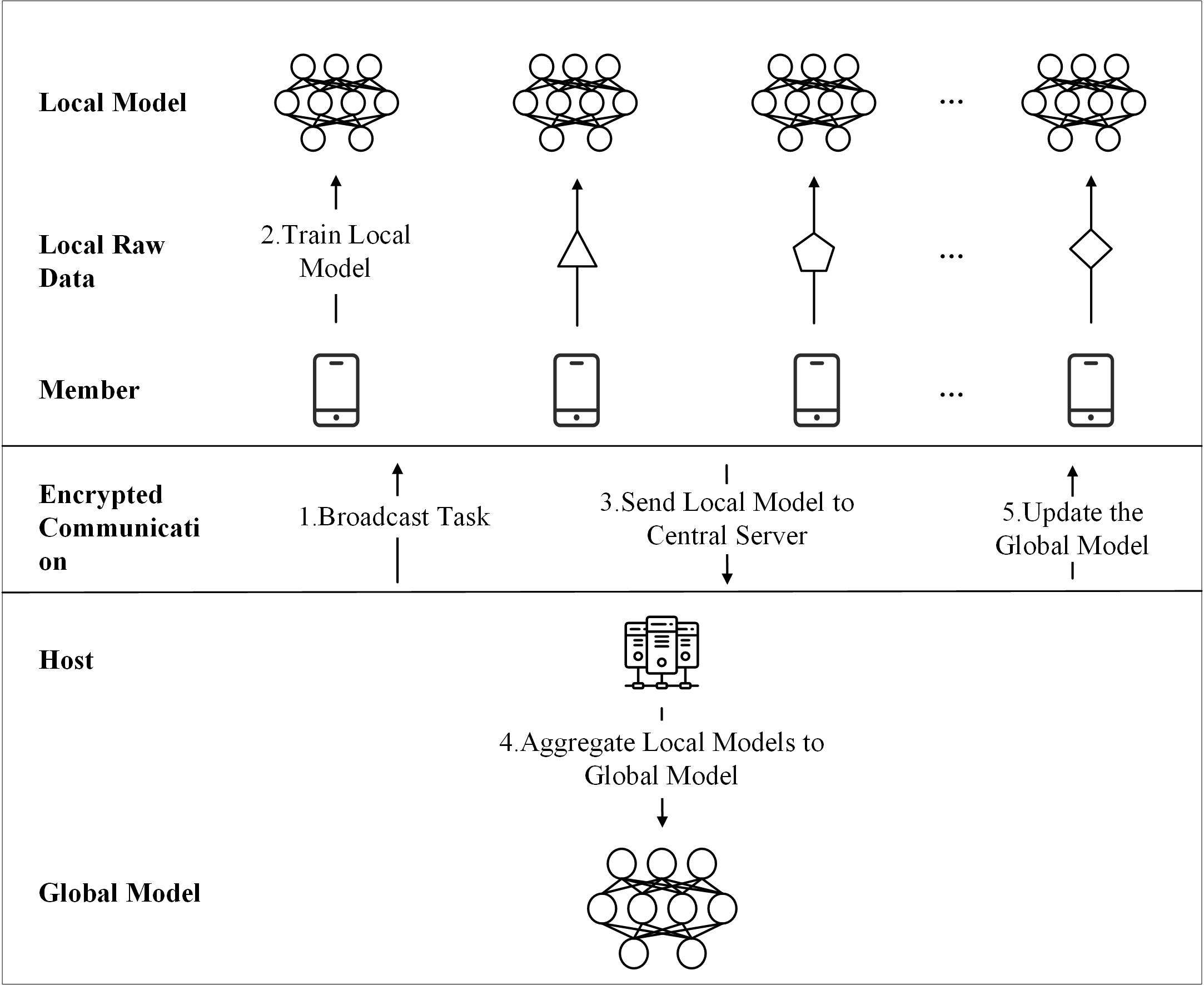}
    \caption{Centralized Cross-device Federated Learning Architecture}
    \label{fig:3}
\end{figure}

In the traditional RS, the server will centrally collect the users' data and construct a RS model by training them together with the item data. Current FL based RS studies \cite{FCF-ammad2019federated}\cite{fedrec-lin2020fedrec}\cite{fedfast-muhammad2020fedfast}\cite{SE-FRS-jalalirad2019simple} consider all the user data as private data, including explicit feedback (rating, comments). However, these data are public and are actively provided to the system by users. Some other user interaction behaviors, including page clicks, favorites, and likes, are also inevitably recorded by the system, as these behaviors are users' active requests on the application. Missing these data will make the RS unable to perform the model initialization (feature engineering, sample labeling), significantly increasing the client device's communication cost and computation load. The personal data that users do not want to disclose is private and need to be protected, such as personal information, personal health information, asset information, work information, and contact information. These data can indeed improve the RS's personalized recommendation capabilities, but these data are sensitive. Once malicious enemies obtain them will bring great distress and risks to the user. 

Therefore, we divide user data into public user data ($Pub\_D$) and private user data ($Pri\_D$).  The PPRSF central server can collect and further compute on the $Pub\_D$. Including the user's typical interaction behavior in the application (Explicit Feedback, Implicit Feedback) and the data that the user actively discloses to the system, which is the essential private user data required by the application context (such as location-based recommender services require user's permission on real-time location data) based on a reliable agreement between the user and the application. $Pri\_D$ is personal data that the user consciously protects, and can only be locally preserved.

From the global perspective, assume N different clients $U_i$ have interaction records with the application, where $i\in [1,N]$. On the client, each client device holds local raw data $D_i$, contains public user data $Pub\_D_i$ and privacy user data $Pri\_D_i$. As the features of local data are basically the same, and the user ID held by each client is different. Such data distribution in FL is called horizontal federated learning \cite{survey-yang2019federated}. And the server holds size of $M$ rows of global item data $Global\_I_j$ as item inventory, where $j\in[1,M]$ , and collected user public data $Pub\_D_i$.

\subsection{Framework Overview}
The PPRSF including 3 essential models within 4-layers hierarchical structure. The system starts with a potentially huge inventory, massive item data is filtered from top to bottom through the layers of Recall, Ranking, and Re-Ranking, and the items that best meet user requirements and interests are presented to users through the client user interface.

\paragraph{Recall Layer - Server}
The Recall layer perform a rough sort on the huge inventory. Input pre-processed $Global\_I_j$ and $Pub\_D_i$ into the Recall model $M_{Recall}(U,I,\Theta)$ , and generate relatively small candidate sets $Recall\_I_{ik}$ for each user $U_i$ that recalled from a large inventory of items, each candidate set contain $K$ rows of item feature data, where $i\in [1,N],j\in [1,M],k\in[1,K]$, and $K<<M$. And then send candidate set to each user's client device. The aim of Recall layer is to reduce the communication cost between server and clients, and the client's computational load during the local model training and inference phase. 
\paragraph{Ranking Layer - Client [1-N]} 
The Ranking layer exists in each client device. Each client device holds local private data $Pri\_D_i$ and an independent local Ranking Model $M_{Rank,i}(U,I,\theta_i)$. Each client $U_i$ receive $Recall\_I_{ik}$ from server, and input into $M_{Rank,i}(U,I,\theta_i)$ together with $Pri\_D_i$, the output is a fine-sorted candidate item list $Rank\_I_{ik}$ accords with preference implied by the local private data, where $i\in[1,N], k\in[1,K]$. The higher the ranking, the higher the relevance to the current user and the Top-T candidate item list $Ranked\_I_{ik}$ will send back to the server for content request, where $k\in [1,T],1<T<K$. 

The local Ranking model training applied the FL paradigm, illustrated in section 3.3. Allows each client to perform the model training and inference on the local device and ensure that the private user data will not be acquired by the host server or any other clients. The Ranking layer aims to extend the performance of personalized recommendations while protecting user privacy.

\paragraph{Re-Rank Layer - Server} 
The Re-Rank Layer on the server will receive the $Ranked\_I_{ik}$ list request, then apply a supplementary strategy (based on application scenarios and system context), consider the diversity, freshness, popularity, and fairness of the final recommendation results before sending it to the client device. When the system encounters a new user or new items, the cold start strategy will perform this layer's task.
\paragraph{Service Layer -Client [1-N]} 
The service layer presents the final recommendation results to each user through the client device's user interface. It is also at this layer that records user interaction behaviors, collects public user data to the server, and saves privacy-related user data on the client device. These records will participate in the next round of model training.

\begin{figure}[ht]
    \centering
    \includegraphics[width=0.90\textwidth]{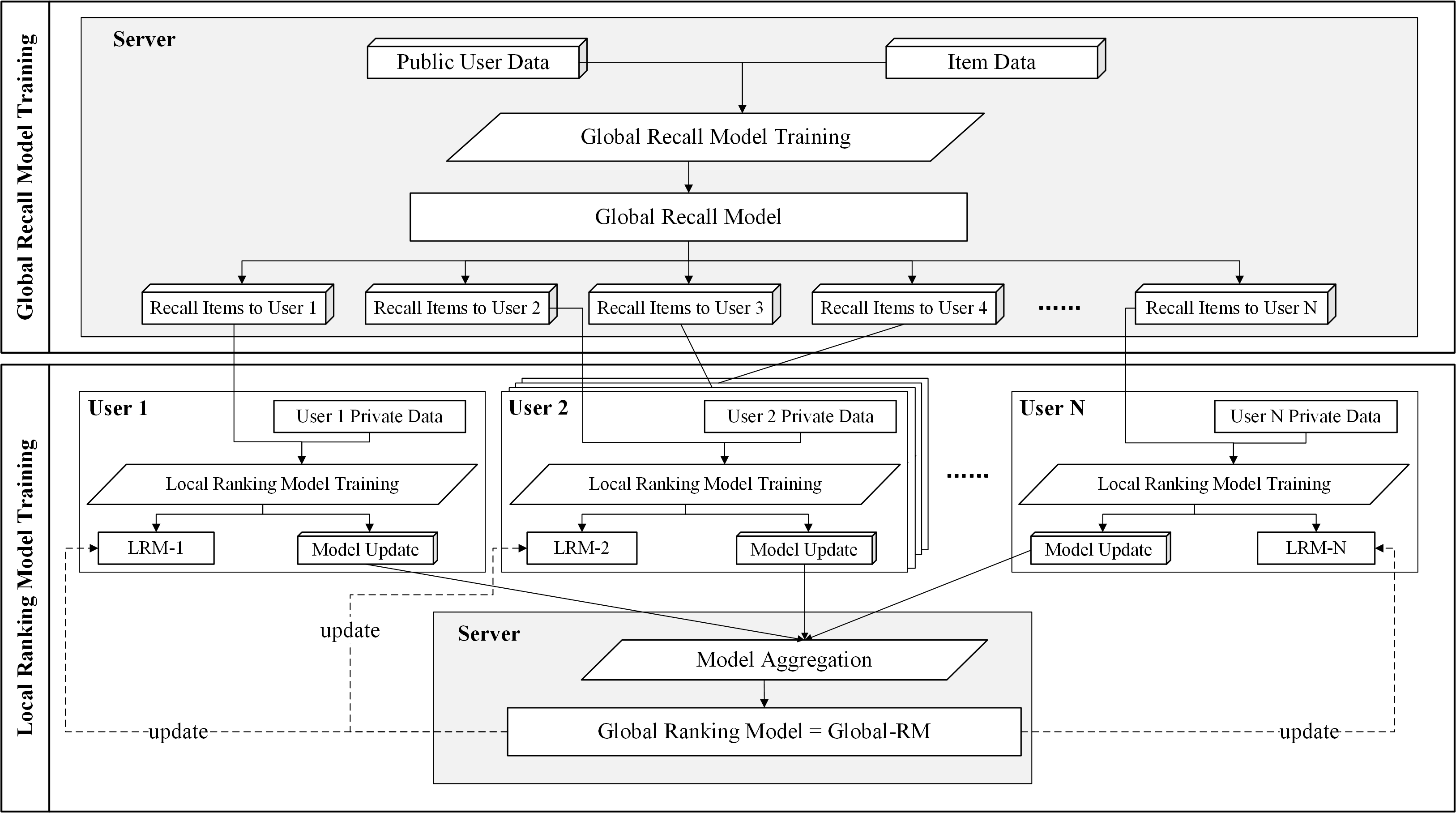} 
    \caption{The Training Process of Global Recall Model and Local Ranking Model in FL-PRS Framework}
    \label{fig:4}
\end{figure}
\subsection{Model Training}
In the PPRSF, the Global Recall Model on the server and the Local Ranking Model on each client are core modules that ensure the effectiveness, feasibility, and privacy of PPRSF. As illustrated in Figure 4, the server performs the training of the Global Recall model; once complete the training work, the output of the model (the recalled item data $Recall\_I_i$) is sent to each client so that the client uses the independent recalled item data and the client's local private user data ($Pri\_D_i$) to train the local ranking model. Subsequently, under the server's cooperation, each client's local models are aggregated on the server to form a global ranking model and then returned to each client for the local model update.

In Section 3.1, based on the definition of private user data and public user data, the server can perform data pre-processing (such as feature engineering, sample labeling, embedding) and jointly modeling on the $Pub\_D_i$ and $Global\_I_j$, where $i\in[1,N],j\in[1,M]$. According to the application scenario and actual data condition, the Global Recall Model can be trained as a content-based model \cite{content-pazzani2007content}, or collaborative filtering based model \cite{collaborative-burke2015robust}, or neural network based model \cite{NCF-he2017neural} \cite{embedding-barkan2016item2vec}. In the absence of private user data, the Global Recall Model can guarantee to a certain extent that the recommendation results conform to the user's preference.
\begin{algorithm}
\caption{Local Ranking Model Training - $Client_i(i \in[1,N]$}
    \begin{algorithmic}[1]
    \REQUIRE $\Theta$,$Recall\_I_{ik}$,$Pri\_D_i$,$f(X,\theta)$
    \ENSURE $\theta_i$,$M_{Rank,i}(U,I,\theta_i)$
    \FOR{each cycle} 
        \STATE Receive Global Parameters $\Theta$ from Server
        \STATE $\theta_i=\Theta$
        \FOR{k=1...K}
            \STATE Sample a batch $b_k$ with $Recall_{ik}$ and $Pri\_D_i$
            \STATE Input $b_k$ into learning algorithm $f(b_k,\theta_i)$
            \STATE Update local model parameter $\theta_i$
        \ENDFOR
        \RETURN $\theta_i$ to Server
    \ENDFOR
    \STATE Local Ranking Model - $M_{Rank,i}(U,I,\theta_i)$
        
    \end{algorithmic}
\end{algorithm}
After the training of Global Recall model is complete, each client is able to receive a filtered item inventory, which reduce the cost of communication. the host server start the Ranking model training task with global model initialization $M_{Rank}(U,I,\Theta)$, then send initialized parameter $\Theta$ and $Recall\_I_i$ to each client. In each client, perform local training work on $Recall\_I_i$ and local private data $Pri\_D_i$, update the local model parameters $\theta_i$, and upload it to the server. As shown in Algorithm 1.

As shown in Algorithm 2, after every client finished their local model training, the host server will receive the model update parameters $\theta_i$ from each client, perform global aggregation to obtain global model parameter $\Theta$, and return it to each client to optimize the local ranking model. After several iterations, each client obtained a global optimal model that able to perform local ranking work. 
\begin{algorithm}
\caption{Global Model Aggregation - Server} 
    \begin{algorithmic}[1]
    \REQUIRE{$\theta_i$}
    \ENSURE{$\Theta$,$M_{Rank}(U,I,\Theta$}
    \FOR{each cycle}
        \STATE Receive $\theta_i$ from clients where $i\in [1,N]$
        \STATE $\Theta = Aggregation_{i=1}^N(\theta_i)$
        \STATE Send $\Theta$ to each client
    \ENDFOR
    \STATE Global Ranking Model - $M_{Rank}(U,I,\Theta)$
    \end{algorithmic}
\end{algorithm}   

\section{Conclusion}
In this paper, we introduced the FL paradigm into the RS and proposed a novel Privacy-preserved RS framework (PPRSF) that enables the training and online inference under the premise of preserving user privacy, satisfying legal and regulatory requirements. This framework can be applied to the application that involving sensitive user information, such as financial service, medical service, and social network.

\bibliographystyle{ieeetr}
\bibliography{references.bib}

\begin{thebibliography}{10}
\providecommand{\url}[1]{#1}
\csname url@samestyle\endcsname
\providecommand{\newblock}{\relax}
\providecommand{\bibinfo}[2]{#2}
\providecommand{\BIBentrySTDinterwordspacing}{\spaceskip=0pt\relax}
\providecommand{\BIBentryALTinterwordstretchfactor}{4}
\providecommand{\BIBentryALTinterwordspacing}{\spaceskip=\fontdimen2\font plus
\BIBentryALTinterwordstretchfactor\fontdimen3\font minus
  \fontdimen4\font\relax}
\providecommand{\BIBforeignlanguage}[2]{{%
\expandafter\ifx\csname l@#1\endcsname\relax
\typeout{** WARNING: IEEEtran.bst: No hyphenation pattern has been}%
\typeout{** loaded for the language `#1'. Using the pattern for}%
\typeout{** the default language instead.}%
\else
\language=\csname l@#1\endcsname
\fi
#2}}
\providecommand{\BIBdecl}{\relax}
\BIBdecl

\bibitem{bobadilla2013recommender}
J.~Bobadilla, F.~Ortega, A.~Hernando, and A.~Guti{\'e}rrez, ``Recommender
  systems survey,'' \emph{Knowledge-based systems}, vol.~46, pp. 109--132,
  2013.

\bibitem{kelly2003implicit}
D.~Kelly and J.~Teevan, ``Implicit feedback for inferring user preference: a
  bibliography,'' in \emph{Acm Sigir Forum}, vol.~37, no.~2.\hskip 1em plus
  0.5em minus 0.4em\relax ACM New York, NY, USA, 2003, pp. 18--28.

\bibitem{zhao2016user}
Z.~Zhao, H.~Lu, D.~Cai, X.~He, and Y.~Zhuang, ``User preference learning for
  online social recommendation,'' \emph{IEEE Transactions on Knowledge and Data
  Engineering}, vol.~28, no.~9, pp. 2522--2534, 2016.

\bibitem{ramakrishnan2001being}
N.~Ramakrishnan, B.~J. Keller, B.~J. Mirza, A.~Y. Grama, and G.~Karypis, ``When
  being weak is brave: Privacy in recommender systems,'' \emph{arXiv preprint
  cs/0105028}, 2001.

\bibitem{zhang2019proactive}
B.~Zhang and S.~S. Sundar, ``Proactive vs. reactive personalization: Can
  customization of privacy enhance user experience?'' \emph{International
  Journal of Human-Computer Studies}, vol. 128, pp. 86--99, 2019.

\bibitem{narayanan2008robust}
A.~Narayanan and V.~Shmatikov, ``Robust de-anonymization of large sparse
  datasets,'' in \emph{2008 IEEE Symposium on Security and Privacy (sp
  2008)}.\hskip 1em plus 0.5em minus 0.4em\relax IEEE, 2008, pp. 111--125.

\bibitem{westin2003social}
A.~F. Westin, ``Social and political dimensions of privacy,'' \emph{Journal of
  social issues}, vol.~59, no.~2, pp. 431--453, 2003.

\bibitem{mcmahan2017communication}
B.~McMahan, E.~Moore, D.~Ramage, S.~Hampson, and B.~A. y~Arcas,
  ``Communication-efficient learning of deep networks from decentralized
  data,'' in \emph{Artificial Intelligence and Statistics}.\hskip 1em plus
  0.5em minus 0.4em\relax PMLR, 2017, pp. 1273--1282.

\bibitem{tran2019federated}
N.~H. Tran, W.~Bao, A.~Zomaya, N.~M. NH, and C.~S. Hong, ``Federated learning
  over wireless networks: Optimization model design and analysis,'' in
  \emph{IEEE INFOCOM 2019-IEEE Conference on Computer Communications}.\hskip
  1em plus 0.5em minus 0.4em\relax IEEE, 2019, pp. 1387--1395.

\bibitem{li2020federated}
T.~Li, A.~K. Sahu, A.~Talwalkar, and V.~Smith, ``Federated learning:
  Challenges, methods, and future directions,'' \emph{IEEE Signal Processing
  Magazine}, vol.~37, no.~3, pp. 50--60, 2020.

\bibitem{brisimi2018federated}
T.~S. Brisimi, R.~Chen, T.~Mela, A.~Olshevsky, I.~C. Paschalidis, and W.~Shi,
  ``Federated learning of predictive models from federated electronic health
  records,'' \emph{International journal of medical informatics}, vol. 112, pp.
  59--67, 2018.

\bibitem{heitmann2010architecture}
B.~Heitmann, J.~G. Kim, A.~Passant, C.~Hayes, and H.-G. Kim, ``An architecture
  for privacy-enabled user profile portability on the web of data,'' in
  \emph{Proceedings of the 1st International workshop on Information
  Heterogeneity and Fusion in Recommender Systems}, 2010, pp. 16--23.

\bibitem{mordacchini2010p2p}
M.~Mordacchini, R.~Baraglia, P.~Dazzi, and L.~Ricci, ``A p2p recommender system
  based on gossip overlays (prego),'' in \emph{2010 10th IEEE International
  Conference on Computer and Information Technology}.\hskip 1em plus 0.5em
  minus 0.4em\relax IEEE, 2010, pp. 83--90.

\bibitem{hecht2012radiommender}
F.~V. Hecht, T.~Bocek, N.~B{\"a}r, R.~Erdin, B.~Kuster, M.~Zeeshan, and
  B.~Stiller, ``Radiommender: P2p on-line radio with a distributed recommender
  system,'' in \emph{2012 IEEE 12th International Conference on Peer-to-Peer
  Computing (P2P)}.\hskip 1em plus 0.5em minus 0.4em\relax IEEE, 2012, pp.
  73--74.

\bibitem{han2004scalable}
P.~Han, B.~Xie, F.~Yang, and R.~Shen, ``A scalable p2p recommender system based
  on distributed collaborative filtering,'' \emph{Expert systems with
  applications}, vol.~27, no.~2, pp. 203--210, 2004.

\bibitem{mcsherry2007mechanism}
F.~McSherry and K.~Talwar, ``Mechanism design via differential privacy,'' in
  \emph{48th Annual IEEE Symposium on Foundations of Computer Science
  (FOCS'07)}.\hskip 1em plus 0.5em minus 0.4em\relax IEEE, 2007, pp. 94--103.

\bibitem{friedman2016differential}
A.~Friedman, S.~Berkovsky, and M.~A. Kaafar, ``A differential privacy framework
  for matrix factorization recommender systems,'' \emph{User Modeling and
  User-Adapted Interaction}, vol.~26, no.~5, pp. 425--458, 2016.

\bibitem{berlioz2015applying}
A.~Berlioz, A.~Friedman, M.~A. Kaafar, R.~Boreli, and S.~Berkovsky, ``Applying
  differential privacy to matrix factorization,'' in \emph{Proceedings of the
  9th ACM Conference on Recommender Systems}, 2015, pp. 107--114.

\bibitem{HE-erkin2012generating}
Z.~Erkin, T.~Veugen, T.~Toft, and R.~L. Lagendijk, ``Generating private
  recommendations efficiently using homomorphic encryption and data packing,''
  \emph{IEEE transactions on information forensics and security}, vol.~7,
  no.~3, pp. 1053--1066, 2012.

\bibitem{HE-gentry2009fully}
C.~Gentry, ``Fully homomorphic encryption using ideal lattices,'' in
  \emph{Proceedings of the forty-first annual ACM symposium on Theory of
  computing}, 2009, pp. 169--178.

\bibitem{mini-collection-granello2004online}
D.~H. Granello and J.~E. Wheaton, ``Online data collection: Strategies for
  research,'' \emph{Journal of Counseling \& Development}, vol.~82, no.~4, pp.
  387--393, 2004.

\bibitem{survey-aledhari2020federated}
M.~Aledhari, R.~Razzak, R.~M. Parizi, and F.~Saeed, ``Federated learning: A
  survey on enabling technologies, protocols, and applications,'' \emph{IEEE
  Access}, vol.~8, pp. 140\,699--140\,725, 2020.

\bibitem{survey-mothukuri2020survey}
V.~Mothukuri, R.~M. Parizi, S.~Pouriyeh, Y.~Huang, A.~Dehghantanha, and
  G.~Srivastava, ``A survey on security and privacy of federated learning,''
  \emph{Future Generation Computer Systems}, 2020.

\bibitem{survey-yang2019federated}
Q.~Yang, Y.~Liu, T.~Chen, and Y.~Tong, ``Federated machine learning: Concept
  and applications,'' \emph{ACM Transactions on Intelligent Systems and
  Technology (TIST)}, vol.~10, no.~2, pp. 1--19, 2019.

\bibitem{DNN-ma2020distributed}
D.~Ma, L.~Li, H.~Ren, D.~Wang, X.~Li, and Z.~Han, ``Distributed rate
  optimization for intelligent reflecting surface with federated learning,'' in
  \emph{2020 IEEE International Conference on Communications Workshops (ICC
  Workshops)}.\hskip 1em plus 0.5em minus 0.4em\relax IEEE, 2020, pp. 1--6.

\bibitem{DNN-ye2020federated}
D.~Ye, R.~Yu, M.~Pan, and Z.~Han, ``Federated learning in vehicular edge
  computing: A selective model aggregation approach,'' \emph{IEEE Access},
  vol.~8, pp. 23\,920--23\,935, 2020.

\bibitem{DNN-zhang2020fedmec}
J.~Zhang, Y.~Zhao, J.~Wang, and B.~Chen, ``Fedmec: improving efficiency of
  differentially private federated learning via mobile edge computing,''
  \emph{Mobile Networks and Applications}, pp. 1--13, 2020.

\bibitem{LR-chen2018privacy}
Y.-R. Chen, A.~Rezapour, and W.-G. Tzeng, ``Privacy-preserving ridge regression
  on distributed data,'' \emph{Information Sciences}, vol. 451, pp. 34--49,
  2018.

\bibitem{LR-nikolaenko2013privacy}
V.~Nikolaenko, U.~Weinsberg, S.~Ioannidis, M.~Joye, D.~Boneh, and N.~Taft,
  ``Privacy-preserving ridge regression on hundreds of millions of records,''
  in \emph{2013 IEEE Symposium on Security and Privacy}.\hskip 1em plus 0.5em
  minus 0.4em\relax IEEE, 2013, pp. 334--348.

\bibitem{LR-sanil2004privacy}
A.~P. Sanil, A.~F. Karr, X.~Lin, and J.~P. Reiter, ``Privacy preserving
  regression modelling via distributed computation,'' in \emph{Proceedings of
  the tenth ACM SIGKDD international conference on Knowledge discovery and data
  mining}, 2004, pp. 677--682.

\bibitem{GBDT-li2020privacy}
Q.~Li, Z.~Wu, Z.~Wen, and B.~He, ``Privacy-preserving gradient boosting
  decision trees,'' in \emph{Proceedings of the AAAI Conference on Artificial
  Intelligence}, vol.~34, no.~01, 2020, pp. 784--791.

\bibitem{GBDT-lu2019differentially}
Y.~Lu, X.~Huang, Y.~Dai, S.~Maharjan, and Y.~Zhang, ``Differentially private
  asynchronous federated learning for mobile edge computing in urban
  informatics,'' \emph{IEEE Transactions on Industrial Informatics}, vol.~16,
  no.~3, pp. 2134--2143, 2019.

\bibitem{GBDT-zhao2018inprivate}
L.~Zhao, L.~Ni, S.~Hu, Y.~Chen, P.~Zhou, F.~Xiao, and L.~Wu, ``Inprivate
  digging: Enabling tree-based distributed data mining with differential
  privacy,'' in \emph{IEEE INFOCOM 2018-IEEE Conference on Computer
  Communications}.\hskip 1em plus 0.5em minus 0.4em\relax IEEE, 2018, pp.
  2087--2095.

\bibitem{DP-triastcyn2019federated}
A.~Triastcyn and B.~Faltings, ``Federated learning with bayesian differential
  privacy,'' in \emph{2019 IEEE International Conference on Big Data (Big
  Data)}.\hskip 1em plus 0.5em minus 0.4em\relax IEEE, 2019, pp. 2587--2596.

\bibitem{DP-wei2020federated}
K.~Wei, J.~Li, M.~Ding, C.~Ma, H.~H. Yang, F.~Farokhi, S.~Jin, T.~Q. Quek, and
  H.~V. Poor, ``Federated learning with differential privacy: Algorithms and
  performance analysis,'' \emph{IEEE Transactions on Information Forensics and
  Security}, 2020.

\bibitem{HE-hao2019efficient}
M.~Hao, H.~Li, X.~Luo, G.~Xu, H.~Yang, and S.~Liu, ``Efficient and
  privacy-enhanced federated learning for industrial artificial intelligence,''
  \emph{IEEE Transactions on Industrial Informatics}, vol.~16, no.~10, pp.
  6532--6542, 2019.

\bibitem{HE-zhang2020batchcrypt}
C.~Zhang, S.~Li, J.~Xia, W.~Wang, F.~Yan, and Y.~Liu, ``Batchcrypt: Efficient
  homomorphic encryption for cross-silo federated learning,'' in \emph{2020
  $\{$USENIX$\}$ Annual Technical Conference ($\{$USENIX$\}$$\{$ATC$\}$ 20)},
  2020, pp. 493--506.

\bibitem{threats-bhagoji2019analyzing}
A.~N. Bhagoji, S.~Chakraborty, P.~Mittal, and S.~Calo, ``Analyzing federated
  learning through an adversarial lens,'' in \emph{International Conference on
  Machine Learning}.\hskip 1em plus 0.5em minus 0.4em\relax PMLR, 2019, pp.
  634--643.

\bibitem{attack-bagdasaryan2020backdoor}
E.~Bagdasaryan, A.~Veit, Y.~Hua, D.~Estrin, and V.~Shmatikov, ``How to backdoor
  federated learning,'' in \emph{International Conference on Artificial
  Intelligence and Statistics}.\hskip 1em plus 0.5em minus 0.4em\relax PMLR,
  2020, pp. 2938--2948.

\bibitem{FCF-ammad2019federated}
M.~Ammad-Ud-Din, E.~Ivannikova, S.~A. Khan, W.~Oyomno, Q.~Fu, K.~E. Tan, and
  A.~Flanagan, ``Federated collaborative filtering for privacy-preserving
  personalized recommendation system,'' \emph{arXiv preprint arXiv:1901.09888},
  2019.

\bibitem{fedrec-lin2020fedrec}
G.~Lin, F.~Liang, W.~Pan, and Z.~Ming, ``Fedrec: Federated recommendation with
  explicit feedback,'' \emph{IEEE Intelligent Systems}, 2020.

\bibitem{fedfast-muhammad2020fedfast}
K.~Muhammad, Q.~Wang, D.~O'Reilly-Morgan, E.~Tragos, B.~Smyth, N.~Hurley,
  J.~Geraci, and A.~Lawlor, ``Fedfast: Going beyond average for faster training
  of federated recommender systems,'' in \emph{Proceedings of the 26th ACM
  SIGKDD International Conference on Knowledge Discovery \& Data Mining}, 2020,
  pp. 1234--1242.

\bibitem{SE-FRS-jalalirad2019simple}
A.~Jalalirad, M.~Scavuzzo, C.~Capota, and M.~Sprague, ``A simple and efficient
  federated recommender system,'' in \emph{Proceedings of the 6th IEEE/ACM
  International Conference on Big Data Computing, Applications and
  Technologies}, 2019, pp. 53--58.

\bibitem{REPTILE-nichol2018first}
A.~Nichol, J.~Achiam, and J.~Schulman, ``On first-order meta-learning
  algorithms,'' \emph{arXiv preprint arXiv:1803.02999}, 2018.

\bibitem{DNN-qi2020privacy}
T.~Qi, F.~Wu, C.~Wu, Y.~Huang, and X.~Xie, ``Privacy-preserving news
  recommendation model training via federated learning,'' \emph{arXiv preprint
  arXiv:2003.09592}, 2020.

\bibitem{privacy-dou2019privacy}
K.~Dou, B.~Guo, and L.~Kuang, ``A privacy-preserving multimedia recommendation
  in the context of social network based on weighted noise injection,''
  \emph{Multimedia Tools and Applications}, vol.~78, no.~19, pp.
  26\,907--26\,926, 2019.

\bibitem{content-pazzani2007content}
M.~J. Pazzani and D.~Billsus, ``Content-based recommendation systems,'' in
  \emph{The adaptive web}.\hskip 1em plus 0.5em minus 0.4em\relax Springer,
  2007, pp. 325--341.

\bibitem{collaborative-burke2015robust}
R.~Burke, M.~P. O’Mahony, and N.~J. Hurley, ``Robust collaborative
  recommendation,'' in \emph{Recommender systems handbook}.\hskip 1em plus
  0.5em minus 0.4em\relax Springer, 2015, pp. 961--995.

\bibitem{NCF-he2017neural}
X.~He, L.~Liao, H.~Zhang, L.~Nie, X.~Hu, and T.-S. Chua, ``Neural collaborative
  filtering,'' in \emph{Proceedings of the 26th international conference on
  world wide web}, 2017, pp. 173--182.

\bibitem{embedding-barkan2016item2vec}
O.~Barkan and N.~Koenigstein, ``Item2vec: neural item embedding for
  collaborative filtering,'' in \emph{2016 IEEE 26th International Workshop on
  Machine Learning for Signal Processing (MLSP)}.\hskip 1em plus 0.5em minus
  0.4em\relax IEEE, 2016, pp. 1--6.

\end{thebibliography}

\end{document}